# Angular distribution of *K*α x rays following nonradiative double electron capture in relativistic collisions of Xe$^{54+}$ ions with Kr and Xe atoms

Bian Yang,[1,2] Deyang Yu,[1,2,*] Konstantin N. Lyashchenko,[3,4] Caojie Shao,[1,2] Zhongwen Wu,[5] Mingwu Zhang,[1,2] Oleg Yu. Andreev,[3] Junliang Liu,[1,2] Zhangyong Song,[1,2] Yingli Xue,[1,2] Wei Wang,[1,2] Fangfang Ruan,[6] Yehong Wu,[7] Rongchun Lu,[1] Chenzhong Dong,[5] and Xiaohong Cai[1]

[1] *Institute of Modern Physics, Chinese Academy of Sciences, Lanzhou 730000, China*
[2] *University of Chinese Academy of Sciences, Beijing 100049, China*
[3] *Department of Physics, St. Petersburg State University, 7/9 Universitetskaya nab., St. Petersburg, 199034, Russia*
[4] *Petersburg Nuclear Physics Institute named by B.P. Konstantinov of National Research Centre "Kurchatov Institute," mkr. Orlova roshcha 1, Gatchina, 188300, Leningrad District, Russia*
[5] *Key Laboratory of Atomic and Molecular Physics & Functional Materials of Gansu Province, College of Physics and Electronic Engineering, Northwest Normal University, Lanzhou 730070, China*
[6] *Department of Medical Imaging, Hangzhou Medical College, Hangzhou 310053, China*
[7] *School of Basic Medical Sciences, Shanxi Medical University, Taiyuan 030001, China*

We present experimental study of nonradiative double electron capture processes in collisions of 95 and 146 MeV/u bare xenon ions with krypton and xenon gaseous atoms at the HIRFL-CSR storage ring. Angular distributions of the characteristic *K*α radiation of the down-charged projectile ions Xe$^{52+*}$ are measured, which are closely related to the magnetic sublevel population of the excited $1s2l_j$ states of Xe$^{52+*}$. It was found that the *K*α$_1$ radiation shows pronounced anisotropic and is sensitive to the collision energies and the target atoms, whereas the *K*α$_2$ radiation gives rise to isotropic. Moreover, obviously difference in the anisotropy parameters of Lyman-α$_1$ of Xe$^{53+*}$ ions and *K*α transitions of Xe$^{52+*}$ ions separately following nonradiative single and double electron capture into the *L*-shell levels of projectiles is obtained and discussed.

## I. INTRODUCTION

The electron capture is one of the fundamental processes in collisions of energetic highly-charged ions with atoms [1,2]. The nonradiative electron capture (NRC) is a process taking place in ion-atom collision in which an electron bound in target atom is captured into a bound state of the projectile ion without radiation emission. The energy and momentum released by the captured electron are shared between the atom and ion [1,2]. The NRC is a dominant charge-transfer mechanism in fast collisions of highly charged high-Z ions with heavy target atoms [2,3], although it usually competes with the radiative electron capture (REC) process [3]. The REC mechanism defines that a weakly bound electron of the target atom transfers into a bound state of the projectile ion accompanied by the emission of a photon for the conservations of energy and momentum [3]. Meanwhile, the nonradiative double electron capture (NRDC) is increasingly important with decreasing projectile energy and increasing target atomic number, as well as with increasing atomic number and charge states of projectile [2,4]. The NRC and NRDC processes have been intensively studied for light and medium mass ions especially in collisions with light atoms for the low-energy region at electron beam ion traps [1,5-7], whereas the experimental investigations in the regime of fast highly charged high-Z ions have been relatively scarce. Recently, heavy ion storage rings [8], such as the Heavy Ion Research Facility at Lanzhou-Cooling Storage Ring (HIRFL-CSR) [9,10] equipped with an internal gas-jet target [11], have provided highly charged heavy-ion beams at projectile energies between tens of and few hundreds MeV/u to extend the studies of the NRC and NRDC processes into the regime of high-Z ions collision with heavy atoms under the single-collision condition.

At present, a few experiments measuring total cross sections and state-selective populations of the NRC and NRDC processes for bare and H-like heavy ions in high-energy region have been reported [2,12-17]. In these studies, special interest has been devoted to the production of excited ionic states and on the measurement of their subsequent radiative decay. The experimental results were separately compared with NRC cross section calculated by the Oppenheimer-Brinkman-Kramers approximation [2], the nonrelativistic continuum distorted-wave method [2], the relativistic eikonal approximation [2,18-20], and relativistic two-center coupled-channel calculations [21,22], as well as NRDC cross sections evaluated from the independent-electron approximation [4,23] and the many-body classical-trajectory Monte Carlo (n-CTMC) calculations [24]. Indeed, the study of the bound-state transitions in heavy ions plays a key role in our understanding of electron-nucleus interaction and electron-electron interaction in the presence of strong Coulomb fields, as well as investigating important information about the relativistic and quantum electrodynamic effects in few-electron high-Z systems [25-29]. Moreover, these observations are important to estimate the charge-state distribution and lifetime of stored ion beams in heavy-ion storage rings [3,8,10,13].

In addition to proper studies of transition energies and probabilities for highly charged heavy ions, measurements of the angular distributions of characteristic x-ray photons provide important information about the structure and dynamics of high-Z ions. One advantage of angle-resolved x-ray studies is that they are more sensitive to the magnetic and retardation effects than the total cross section obtained as a result of integration over the emission angles [3,13,30-32]. In previous works [12,33], we have the first time addressed angular distribution of the subsequent Lyman-$\alpha_1$ radiation of H-like heavy ions produced by the NRC process in fast collisions of bare Xe ions with Kr and Xe atoms. The negative anisotropy parameters presented a strong population of the magnetic sublevels with $m_j = \pm 1/2$ for the $2p_{3/2}$ state of $Xe^{53+*}$ ions at projectile energies of 95 and 146 MeV/u. Therefore, the results lead to emission of strongly polarized Lyman-$\alpha_1$ photons. The obtained results show weak target dependence and significant energy dependence. Namely, if the projectile energy increases to 197 MeV/u, the Lyman-$\alpha_1$ radiation for the Kr target becomes much less anisotropic and the magnetic sublevels are almost equally populated. Further, the experimental findings are compared with the results of the corresponding REC mechanism (since there are no detailed computations available for cross section related to magnetic sublevels of the NRC process to the best of our knowledge), obviously different energy dependence of the magnetic-sublevel population is obtained. Moreover, the measured total intensity ratios of the Lyman-$\alpha_1$ and Lyman-α$_2$(+$M$1) transitions reveal that target electrons are preferentially captured to the $2p_{1/2}$ and $2s_{1/2}$ subshells rather than the $2p_{3/2}$ one in fast collisions of high-Z ions with atoms. These results might serve as a test for the theory of NRC dynamics and help reveal the population mechanism of excited states in fast collisions of highly charged high-Z ions with atoms.

While these previous studies on NRC have provided first insights into the magnetic sublevel population of high-Z hydrogen-like ions [12,33], the analogous information for the NRDC process remains entirely unexplored. To date, there are no reliable theoretical calculations or experimental data on the magnetic sublevel population following NRDC with bare high-Z ions in relativistic collisions. This lack of knowledge represents a significant gap, as NRDC becomes increasingly important for heavy targets and lower energies, and its dynamics inherently involve two-electron correlations and their interplay with the strong Coulomb field. Angular distribution measurements of the subsequent characteristic x rays, such as the $K\alpha$ radiation, provide a direct pathway to access this information, since the anisotropy parameters are sensitive to the magnetic sublevel population of the excited states formed in the capture process.

Therefore, the primary goal of this work is to perform the first experimental study of the angular distributions of $K\alpha_1$ and $K\alpha_2$ x rays following NRDC. We present measurements for collisions of 95 and 146 MeV/u bare Xe$^{54+}$ ions with Kr and Xe gas targets at the HIRFL-CSR storage ring. From these distributions, we extract the anisotropy parameters, which are closely related to the magnetic sublevel population of the excited He-like Xe$^{52+*}$ $1s2l_j$ states. Furthermore, we compare these results for the NRDC process with the well-studied anisotropy of the Lyman-α$_1$ radiation following NRC, highlighting fundamental differences in the population mechanisms between single- and double-electron transfer in relativistic heavy-ion collisions. In the next section, the experiment and x-ray spectra are described. The results and discussion are presented in Sec. III. Finally, the conclusion is given in Sec. IV.

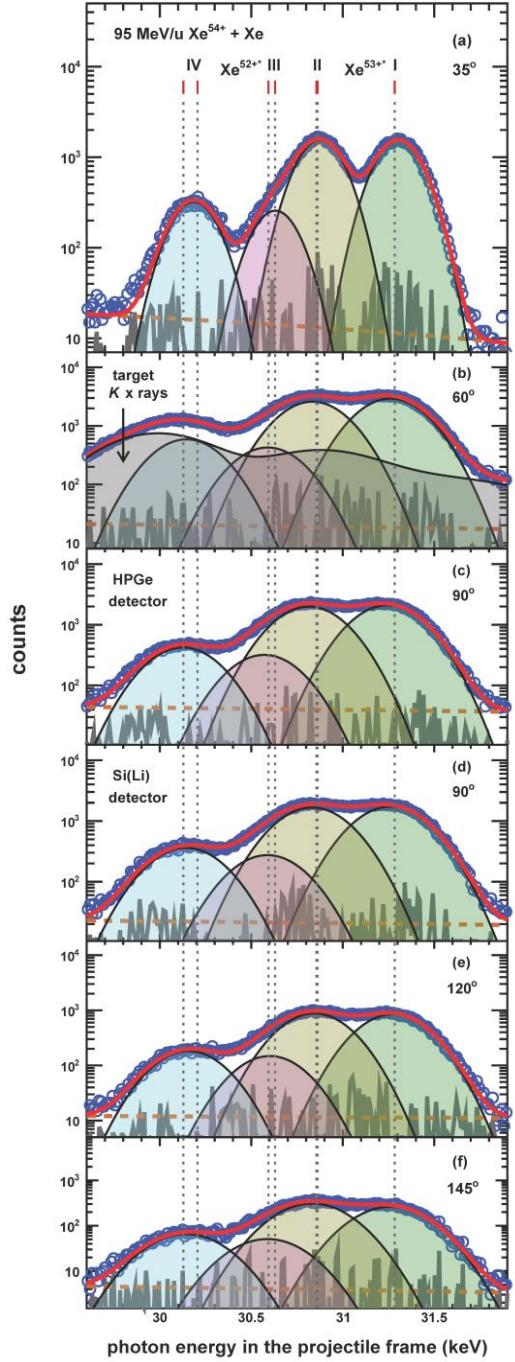

FIG. 1. X-ray spectra of the $K$-shell radiation from $Xe^{53+*}$ and $Xe^{52+*}$ produced by electron capture in collisions of 95 MeV/u $Xe^{54+}$ with Xe at observation angles of (a) 35° recorded with a HPGe−GLP762 detector, (b) 60° with a HPGe-GLP758 detector, (c) 90° with a HPGe-GLP756 detector, (d) 90° with a Si(Li)-SLP198B detector, (e) 120° with a HPGe-GLP749 detector, and (f) 145° with a Si(Li)-SLP7666 detector. The peaks marked as I and II correspond to the transitions of Lyman-$\alpha_1$ ($2p_{3/2} \rightarrow 1s_{1/2}$) and Lyman-$\alpha_2$(+$M1$) ($2p_{1/2}, 2s_{1/2} \rightarrow 1s_{1/2}$) of the down-charged ions $Xe^{53+*}$, respectively. The peaks III and IV are the radiations of $K\alpha_1$ $\left((1s_{1/2}2p_{3/2})_1, (1s_{1/2}2p_{3/2})_2 \rightarrow 1s^2_{1/2}\right)$ and $K\alpha_2$ $\left((1s_{1/2}2p_{1/2})_1, (1s_{1/2}2s_{1/2})_1 \rightarrow 1s^2_{1/2}\right)$ from the doubly down-charged ions $Xe^{52+*}$, respectively. The energies of these transitions [34,35] are illustrated by dashed vertical lines. Intensities of these peaks are determined by a fitting procedure of four Gaussian functions with a linear background. The fitting residuals are also shown. Determination of transition intensities when x rays emitted from projectiles overlap with the *K* x rays resulting from target ionization at 60° was described in the previous paper [36].

## II. EXPERIMENT AND X-RAY SPECTRA

The experiments were carried out using the internal gas-jet target [11] in the experimental ring of the HIRFL-CSR storage ring [9,10]. The detailed description of the experimental arrangement can be found in Refs. [12,36]. In brief, the number of bare xenon ions at an energy of 95 or 146 MeV/u per injection was on the order of $10^7$. An efficient electron cooling provided beams with very low emittance of $1\,\pi\text{mm}\cdot\text{mrad}$ and a longitudinal momentum spread of $\Delta p/p \sim 10^{-5}$. After the injection and the following electron cooling, the ion beams interacted with the target jets of Kr or Xe with typical densities between $10^{12}$ and $10^{13}$ particles/cm$^2$. X rays emitted from the interaction region were detected by four high-purity germanium (HPGe) detectors and two lithium-drifted silicon [Si(Li)] detectors, covering observations angles in the range between 35° and 145° with respect to the beam axis. The typical energy resolution (i.e., the full width at half maximum) of the present detectors was about 300 eV at around 30 keV [36].

As an example, Figure 1 shows typical projectile x-ray spectra obtained in the collisions of 95 MeV/u Xe$^{54+}$ with Xe atoms, at observation angles of 35°, 60°, 90°, 120°, and 145° with respect to the beam axis. Note that the photon energies with respect to the transition energies in the projectile frame have been Doppler-corrected from the laboratory frame. The characteristic transitions arising from single capture (Lyman-α radiation in singly down-charged ions Xe$^{53+*}$) and double capture (*K*α transitions in doubly down-charged ions Xe$^{52+*}$) are clearly visible, besides, intensities of the Lyman-α spectral lines are much bigger than that of the *K*α lines. The transition energies of Lyman-α$_1$ ($2p_{3/2} \to 1s_{1/2}$), Lyman-α$_2$ ($2p_{1/2} \to 1s_{1/2}$), and *M*1 ($2s_{1/2} \to 1s_{1/2}$) of Xe$^{53+*}$ ions are 31.284 keV, 30.856 keV, and 30.863 keV [35]. Also the transition energies of *M*1 ($(1s_{1/2}2s_{1/2})_1 \to 1s^2$), *K*α$_2$ ($(1s_{1/2}2p_{1/2})_1 \to 1s^2$), *M*2 ($(1s_{1/2}2p_{3/2})_2 \to 1s^2$), and *K*α$_1$ ($(1s_{1/2}2p_{3/2})_1 \to 1s^2$) of Xe$^{52+*}$ ions are 30.129 keV, 30.206 keV, 30.594 keV, and 30.630 keV, respectively [34]. Here, the Lyman-α$_1$ of $2p_{3/2} \to 1s_{1/2}$ transition was marked as I. The lines of Lyman-α$_2$(+*M*1) include $2p_{1/2}, 2s_{1/2} \to 1s_{1/2}$ transitions marked as II. *K*α$_1$(+*M*2) corresponding to $(1s_{1/2}2p_{3/2})_1, (1s_{1/2}2p_{3/2})_2 \to 1s^2$ transitions marked as III and *K*α$_2$(+*M*1) corresponding to $(1s_{1/2}2p_{1/2})_1, (1s_{1/2}2s_{1/2})_1 \to 1s^2$ transitions marked as IV. Because of additional broadening introduced by the Doppler effect of fast projectiles and limited energy resolution of the photon detectors used in the present experiment, these pairs of transitions cannot be resolved by photon detectors used in the present work. Recently, it is reported that cryogenic calorimeter detector can resolve *K*α$_1$, *M*2, *K*α$_2$, and *M*1 transitions of Xe$^{52+*}$ with very small energy differences in 50 MeV/u Xe$^{54+}$+Xe collisions [37], as well as the ones of U$^{90+*}$ in collisions of 10.225 MeV/u U$^{91+*}$ ions with electrons [38]. At each observation angle, the number of counts (i.e., intensities) recorded in the Lyman-α$_1$, Lyman-α$_2$(+*M*1), *K*α$_1$(+*M*2), and *K*α$_2$(+*M*1) spectral lines were obtained by fitting the corresponding peaks in the spectra with four Gaussian functions and a linear background (as described in detail in Refs. [12,36]), and then corrected for the energy-dependent detector efficiency.

## III. RESULTS AND DISCUSSION

The measured x-ray spectra in the present experiment indicate that the Lyman transitions of $Xe^{53+*}$ belong to the NRC mechanism due to the almost absence of REC photon emission in the $K$- and $L$-shells of the projectile. Another reason is that NRC cross sections calculated with the REA method [18-20] are much larger than that of REC based on nonrelativistic dipole approximation [3], as discussed in our previous studies [12,36]. For the case of the $K\alpha_1(+M2)$ and $K\alpha_2(+M1)$ transitions of $Xe^{52+*}$, the double electron capture is resulting from NRDC, as described in the previous paper [17]. The double radiative electron capture, the radiative double-electron capture, and double capture with one REC and one NRC are not important and can be ignored [17].

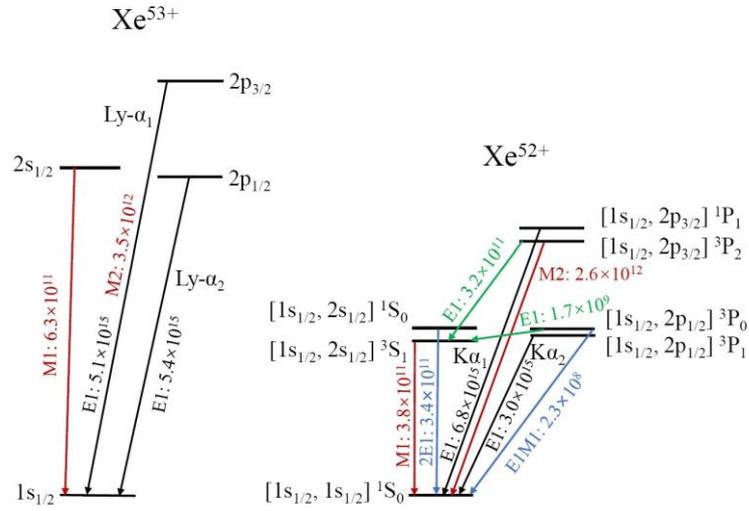

FIG. 2. Level scheme and main decay modes of the $L$-shell excited states for H-like and He-like xenon ions. The $K$- and $L$-shell levels of H-like and He-like xenon along with the transitions contributing to the observed Lyman-$\alpha_1$, Lyman-$\alpha_2(+M1)$, $K\alpha_1(+M2)$, and $K\alpha_2(+M1)$ lines. The corresponding multipolarities and transition rates (calculated by using the flexible atomic code [39]) are indicated and given in units of s$^{-1}$ separately.

Generally, the angular distribution of the $K\alpha_1(+M2)$ or $K\alpha_2(+M1)$ lines can be obtained by correcting their intensities for the solid angles covered by the corresponding photon detectors. This correction introduces additional systematic uncertainties [3,40]. In our experiment, we could benefit from the fact that the Lyman-$\alpha_2(+M1)$ transition arising from the decay of the $2p_{1/2}$ and $2s_{1/2}$ levels is known to be the isotropic distribution [3,41]. Consequently, the Lyman-$\alpha_2(+M1)$ line provides an ideal tool to measure a possible anisotropy of energetically close $K\alpha$ or Lyman-$\alpha_1$ transitions. By using this transition for normalization purpose, various systematic uncertainties (associated for example with corrections for solid angle of detectors, the difference of detection efficiencies) are substantially reduced or cancel out. This method has been used in previous studies for the angular distributions of the Lyman-$\alpha_1$ radiation originating from REC [32], NRC [12,33], excitation [42,43], and simultaneous excitation and ionization processes [44] in fast collisions of bare, H-like, and He-like heavy ions. Similarly, angular distributions for the $K\alpha_1$ and $K\alpha_2$ lines resulting from REC [45-47], resonant transfer and excitation into excited states of initially H-like heavy ions [48], and excitation of initially He-like heavy ions [40,42,49] in relativistic collisions can be obtained separately by intensity ratio between the corresponding transition and the Lyman-$\alpha_2$ (+$M$1) line for different observation angles.

In the following, we focus on the angular distributions of characteristic $K\alpha$ x rays following the NRDC of bare projectiles. Such an angle-resolved study could enable us to gain insight into a mechanism of formation of excited ionic substates of two-electron heavy ions in high-energy collisions of high-Z ions with multielectron atoms. The information on the magnetic sublevel population of the excited few-electron high-Z ions can be directly extracted from the angular distribution of deexcitation photons.

By definition, the intensity of the $K\alpha$ transition (i.e., $K\alpha_1$(+$M$2) or $K\alpha_2$(+$M$1)) in He-like xenon as a function of the observation angle can be generally described by the following formula [40,45,50]

$$\frac{d\sigma_{K\alpha}}{d\Omega}(\theta_{\text{lab}}) = \frac{\sigma_{K\alpha}^{\text{total}}}{4\pi} \frac{1-\beta^2}{(1-\beta\cos\theta_{\text{lab}})^2} \times \left\{1 + \beta_{20}^{K\alpha}\left[1 - \frac{3}{2}\frac{(1-\beta^2)\sin^2\theta_{\text{lab}}}{(1-\beta\cos\theta_{\text{lab}})^2}\right]\right\}, \quad (1)$$

where $\theta_{\text{lab}}$ is the photon observation angle with respect to the beam direction in the laboratory frame, $\beta$ is the reduced projectile velocity in the unit of light speed, and $\sigma_{K\alpha}^{\text{total}}$ denotes the total cross section of the $K\alpha$ transition (i.e., the cross section at the angle for which the angle-dependent part of Eq. (1) is zero). $\beta_{20}^{K\alpha}$ is the anisotropy parameter of the corresponding $K\alpha$ transition. The factor $\frac{1-\beta^2}{(1-\beta\cos\theta_{\text{lab}})^2}$ describes the (relativistic) solid angles transformation between the projectile and laboratory frames, it can be canceled out for the intensities ratio of transition lines measured at the same beam energy.

Considering the angular distribution of the Lyman-$\alpha_2$(+$M$1) transition in the laboratory frame can be described as [12,50]

$$\frac{d\sigma_{\text{Ly-}\alpha_2(+M1)}}{d\Omega}(\theta_{\text{lab}}) = \frac{\sigma_{\text{Ly-}\alpha_2(+M1)}^{\text{total}}}{4\pi}\frac{1-\beta^2}{(1-\beta\cos\theta_{\text{lab}})^2}. \quad (2)$$

Therefore, the angular distribution of the intensity ratio between the $K\alpha$ and Lyman-$\alpha_2$(+$M$1) transitions (following separately NRDC and NRC in the present work) can be represented by

$$\frac{I_{K\alpha}(\theta_{\text{lab}})}{I_{\text{Ly-}\alpha_2(+M1)}(\theta_{\text{lab}})} = \frac{\sigma_{K\alpha}^{\text{total}}}{\sigma_{\text{Ly-}\alpha_2(+M1)}^{\text{total}}} \times \left\{1 + \beta_{20}^{K\alpha}\left[1 - \frac{3}{2}\frac{(1-\beta^2)\sin^2\theta_{\text{lab}}}{(1-\beta\cos\theta_{\text{lab}})^2}\right]\right\}. \quad (3)$$

The angular emission patterns of the $K\alpha_1$(+$M$2) or $K\alpha_2$(+$M$1) spectral lines are determined by the corresponding anisotropy parameter $\beta_{20}^{K\alpha_1(+M2)}$ or $\beta_{20}^{K\alpha_2(+M1)}$ which is in turn related to the alignment of the corresponding magnetic sublevels whose exact form depends on the transition under consideration.

For angular distributions of ground-state radiation from the $[1s_{1/2}, 2p_{3/2}]\,^1P_1$, $[1s_{1/2}, 2p_{3/2}]\,^3P_2$, $[1s_{1/2}, 2p_{1/2}]\,^3P_1$, $[1s_{1/2}, 2s_{1/2}]\,^3S_1$ states of He-like xenon produced by double electron capture, we use only the anisotropy parameter $\beta_{20}$. For the decay of the state $[1s_{1/2}, 2p_{3/2}]\,^3P_2$, corresponding magnetic sublevel population is described by anisotropy parameters $\beta_{20}$ and $\beta_{40}$. That is, Eq. (3) for the $K\alpha_1$(+$M$2) transition has to be modified to account for an additional term proportional to the $\beta_{40}$ [3,40,51]. However, at present there is not applicable relativistic calculations of magnetic-sublevel cross sections for the NRDC process to the best of our knowledge. The corresponding anisotropy parameters $\beta_{20}$ and $\beta_{40}$ for the specified bound-bound electron transition could not be deduced for comparation. In addition, the alignment parameter $\mathcal{A}_{40}$ does not affect the emission of the dipole radiation, and its contribution can hardly be observed at leading electric dipole transitions. Moreover, we found that the $\beta_{40}$ parameters were very small and can be neglected in analysis of angular distribution of the $[1s_{1/2}, 2p_{3/2}]\ ^3P_2$ radiation producing from REC into H-like heavy projectile ions [45,51] and from excitation of He-like heavy projectile ions [40] for projectile energies ranging from a few tens to a few hundreds of MeV/u. Therefore, we do not consider $\beta_{40}$ for the state $[1s_{1/2}, 2p_{3/2}]\ ^3P_2$ and use Eq. (3) for all cases in the present work.

For analyzing experimental results of the intensity ratios between $K\alpha_1(+M2)$ or $K\alpha_2(+M1)$ and Lyman-$\alpha_2(+M1)$ transitions contributing to the respective spectral lines, the decay schemes of H-like and He-like xenon have to be taken into account. As can be seen in Fig. 2, six excited states of He-like xenon may be populated and give contribution to the subsequent radiative decay for double electron capture into bare heavy ions. $[1s_{1/2}, 2p_{3/2}]\ ^3P_2$ and $[1s_{1/2}, 2p_{1/2}]\ ^3P_0$ states have two decay branches each. As a result, the electron-capture into the $[1s_{1/2}, 2p_{3/2}]\ ^3P_2$ state contributes to both $K\alpha_1$ and $K\alpha_2$ lines, and the $[1s_{1/2}, 2p_{1/2}]\ ^3P_0$ state contributes to the $K\alpha_2$ lines via the $E1$ decay branch to the $[1s_{1/2}, 2s_{1/2}]\ ^3S_1$ state. The direct decay from the $[1s_{1/2}, 2p_{1/2}]\ ^3P_0$ state to the ground state proceeds via two-photon $E1M1$ emission and thus it does not contribute to the $K\alpha$ lines Similarly, the $[1s_{1/2}, 2s_{1/2}]\ ^1S_0$ state does not contribute due to its two-photon $2E1$ decay.

For the Lyman-$\alpha_1$ decay of H-like heavy ions, the corresponding anisotropy parameter $\beta_{20}$ includes the alignment parameter $\mathcal{A}_{20}$ and the structure function $f\ (E1, M2)$ [3,12]. This function describes the multipole mixing (interference) between the leading electric dipole transition $E1$ and the much weak magnetic quadrupole $M2$ decay [3,30]. The multipole mixing leads to a 10% enhancement of the anisotropy of the Lyman-$\alpha_1$ radiation for the hydrogenlike xenon ion [52].

In contrast to the Lyman-$\alpha_1$ transition, no multipole mixing can occur for the decay of $[1s_{1/2}, 2p_{1/2}]\ ^3P_1$ and $[1s_{1/2}, 2p_{3/2}]\ ^1P_1$ states of He-like xenon ions (only decay via fast $E1$ transition to the ground state), as well as $[1s_{1/2}, 2p_{3/2}]\ ^3P_2$ (decay via $M2$ transition to the ground state) and $[1s_{1/2}, 2s_{1/2}]\ ^3S_1$ (decay via $M1$ transition to the ground state) states [42,45], as displayed in Fig. 2. For the transitions of $^3P_1 \rightarrow\ ^1S_0$, $^1P_1 \rightarrow\ ^1S_0$, and $^3S_1 \rightarrow\ ^1S_0$, their angular distributions (i.e. emission patterns) follow Eq. (1) with the anisotropy parameter $\beta_{20} = \frac{1}{\sqrt{2}}\mathcal{A}_{20}$, where the corresponding alignment parameter reads $\mathcal{A}_{20} = \sqrt{2}\frac{\sigma(1,\pm1)-\sigma(1,0)}{\sigma(1,\pm1)+\sigma(1,0)}$ [42,45]. Here, the partial cross sections $\sigma(J, M)$ denote the population of the magnetic sublevel of angular momentum $J$ and its projectile $M$ (on the quantization axis), and could describe the dynamics (or explore the properties) of NRDC into the bare projectiles [53]. For the $^3P_2 \rightarrow\ ^1S_0$ transition, the corresponding parameters are defined by $\beta_{20} = -\sqrt{\frac{5}{14}}\mathcal{A}_{20}$ and $\mathcal{A}_{20} = -\sqrt{\frac{10}{7}}\frac{\sigma(2,0)+\sigma(2,\pm1)-2\sigma(2,\pm2)}{\sigma(2,0)+2\sigma(2,\pm1)+2\sigma(2,\pm2)}$ [45].

If the fine-structure components of the $K\alpha_1(+M2)$ and $K\alpha_2(+M1)$ radiation could be resolved experimentally, anisotropic behavior of these four transitions could be studied and provide information on the (individual) anisotropy parameters $\beta_{20}$. However, since the splitting between the $^3P_2$ and $^1P_1$ levels is below the energy resolution of used HPGe and Si(Li) x-ray detectors, only an incoherent superposition of the $^1P_1 \rightarrow\ ^1S_0$ and $^3P_2 \rightarrow\ ^1S_0$ radiation has been observed in the current experiments. Therefore, the angular distribution of the overall $K\alpha_1(+M2)$ radiation is completely determined by the angular distributions of the $E1$ $^1P_1 \to {}^1S_0$ and $M2$ $^3P_2 \to {}^1S_0$ transitions, taken with nonstatistical weights as determined by the NRDC (formation) mechanism of the two excited levels $^3P_2$ and $^1P_1$:

$$W_{K\alpha_1(+M2)}(\theta) = N(^1P_1)W_{E1}(\theta) + N(^3P_2)W_{M2}(\theta) \sim 1 + \beta_{20}^{K\alpha_1(+M2)}(^{1,3}P_{J=1,2} \to {}^1S_0)P_2(\cos\theta). \quad (4)$$

Apparently, the angular dependence of the $K\alpha_1(+M2)$ radiation of He-like ions follows the typical $\sim 1 + \beta_{20} \cdot P_2(\cos\theta)$ shape, but with an overall anisotropy parameter is defined by the following formula [40,45]

$$\beta_{20}^{K\alpha_1(+M2)}(^{1,3}P_{J=1,2} \to {}^1S_0) = \frac{1}{\sqrt{2}}N(^1P_1)\mathcal{A}_{20}(^1P_1) - \sqrt{\frac{5}{14}}N(^3P_2)\mathcal{A}_{20}(^3P_2). \quad (5)$$

Here, the $\beta_{20}^{K\alpha_1(+M2)}$ for the $K\alpha_1(+M2)$ radiation depends on both the alignment parameters $\mathcal{A}_{20}(^1P_1)$ and $\mathcal{A}_{20}(^3P_2)$ of the $[1s_{1/2}, 2p_{3/2}]\,^1P_1$ and $[1s_{1/2}, 2p_{3/2}]\,^3P_2$ states and their weight factors $N(^1P_1)$ and $N(^3P_2)$, respectively. These weights describe the contribution of the individual $^1P_1 \to {}^1S_0$ and $^3P_2 \to {}^1S_0$ transitions to the overall $K\alpha_1(+M2)$ and are given by the (relative) population of the $^3P_2$ and $^1P_1$ levels, with $N(^1P_1) + N(^3P_2) = 1$.

Similarly, the $\beta_{20}^{K\alpha_2(+M1)}$ for the $K\alpha_2(+M1)$ radiation is given by the following formula [40]

$$\beta_{20}^{K\alpha_2(+M1)}(^3S, P_{J=1} \to {}^1S_0) = \frac{1}{\sqrt{2}}N(^3S_1)\mathcal{A}_{20}(^3S_1) + \frac{1}{\sqrt{2}}N(^3P_1)\mathcal{A}_{20}(^3P_1), \quad (6)$$

where $\mathcal{A}_{20}(^3S_1)$ and $\mathcal{A}_{20}(^3P_1)$ are alignment parameters of the $[1s_{1/2}, 2s_{1/2}]\,^3S_1$ and $[1s_{1/2}, 2p_{1/2}]\,^3P_1$ states, respectively. $N(^3S_1)$ and $N(^3P_1)$ are their relative populations, with $N(^3S_1) + N(^3P_1) = 1$. Here, for the He-like xenon produced by the NRDC process, all four levels of $[1s_{1/2}, 2p_{3/2}]\,^1P_1$, $[1s_{1/2}, 2p_{3/2}]\,^3P_2$, $[1s_{1/2}, 2p_{1/2}]\,^3P_1$, $[1s_{1/2}, 2s_{1/2}]\,^3S_1$ can be in principle populated and thus contribute to the $K\alpha_1(+M2)$ and $K\alpha_2(+M1)$ transitions. Therefore, in contrast to the H-like case, it is not possible to extract directly the alignment of one of these states by the fit of the Eq. (3) to the angular distribution of the intensity ratios between the $K\alpha_1(+M2)$ or $K\alpha_2(+M1)$ and Lyman-$\alpha_2(+M1)$ transitions. Moreover, the results in the present work are different from the case of alignment parameters $\mathcal{A}_{20}$ of the $[1s_{1/2}, 2p_{3/2}]\,^1P_1$ and $[1s_{1/2}, 2p_{3/2}]\,^3P_2$ states produced by the REC process with H-like high-Z projectile, which could be expressed in terms of the corresponding alignment parameter $\mathcal{A}_{20}(2p_{3/2})$ from REC with bare projectiles by using the independent particle model [46].

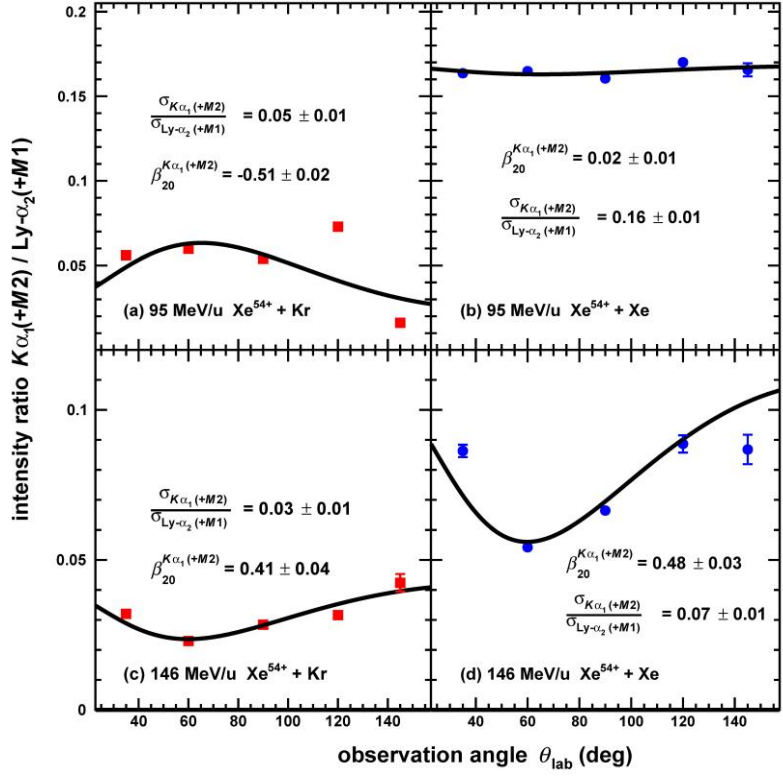

FIG. 3. Angular distributions of the intensity ratios between the $K\alpha_1(+M2)$ and Lyman-$\alpha_2(+M1)$ transitions for He-like and H-like xenon ions produced by the nonradiative double and single electron capture for 95 and 146 MeV/u $Xe^{54+}$ collisions with Kr and Xe targets. The solid lines are fittings of the experimental data according to Eq. (3), the anisotropy parameters $\beta_{20}^{K\alpha_1(+M2)}$ and total cross-section ratios $\sigma_{K\alpha_1(+M2)}^{\text{total}}/\sigma_{\text{Ly-}\alpha_2(+M1)}^{\text{total}}$ are also displayed. The error bars are estimated by the statistics of data.

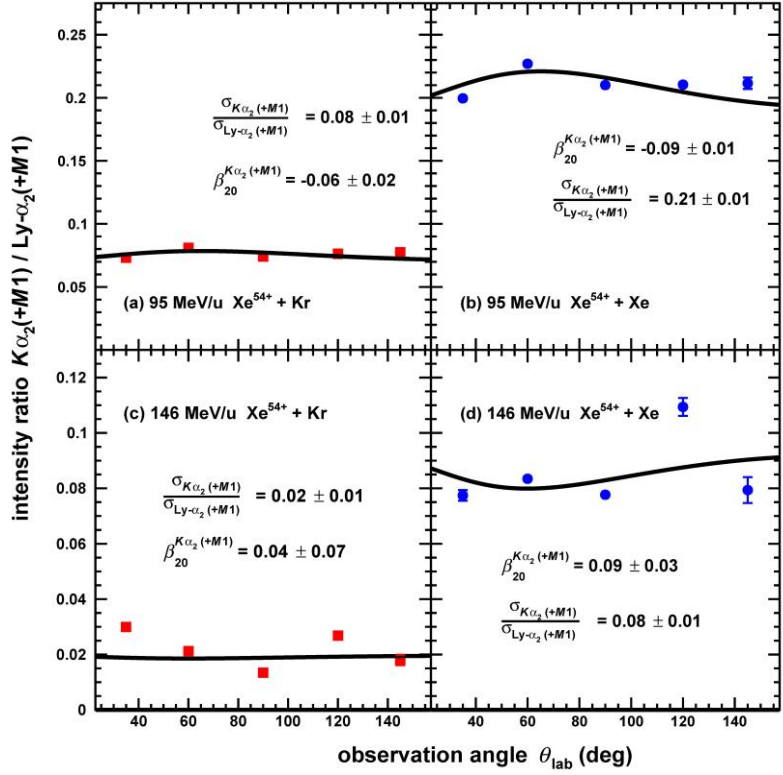

FIG. 4. Angular distributions of the intensity ratios between the $K\alpha_2(+M1)$ and Lyman-$\alpha_2(+M1)$ transitions of He-like and H-like xenon ions produced by the nonradiative double and single electron capture for 95 and 146 MeV/u $Xe^{54+}$ collisions with Kr and Xe targets. The solid lines are fittings of the experimental data according to Eq. (3), the anisotropy parameters $\beta_{20}^{K\alpha_2(+M1)}$ and total cross-section ratios $\sigma_{K\alpha_2(+M1)}^{total}/\sigma_{Ly\text{-}\alpha_2(+M1)}^{total}$ are also displayed. The error bars are estimated by the statistics of data.

In order to obtain the experimental results of the anisotropy parameters $\beta_{20}^{K\alpha}$ and the total cross-section ratios $\sigma_{K\alpha}^{total}/\sigma_{Ly\text{-}\alpha_2(+M1)}^{total}$ for He-like and H-like xenon produced by the nonradiative double and single electron capture in collisions of 95 and 146 MeV/u $Xe^{54+}$ ions with Kr and Xe atoms, we fitted angular distributions of the intensity ratios between the $K\alpha$ and Lyman-$\alpha_2(+M1)$ transitions with Eq. (3) allowing $\sigma_{K\alpha}^{total}/\sigma_{Ly\text{-}\alpha_2(+M1)}^{total}$ and $\beta_{20}^{K\alpha}$ to be free parameters. Figure 3 shows angular distributions of the intensity ratios between $K\alpha_1(+M2)$ spectral lines of $Xe^{52+*}$ and Lyman-$\alpha_2(+M1)$ lines of $Xe^{53+*}$, as well as the corresponding fitting results in collisions of 95 and 146 MeV/u $Xe^{54+}$ projectiles with Kr and Xe targets. The results for the intensity ratios between the $K\alpha_2(+M1)$ and Lyman-$\alpha_2(+M1)$ transitions are displayed in Fig. 4. The deduced anisotropy parameters and the total cross-section ratios were then collected in Figs. 6 and 7 for comparison. The uncertainties of experimental results displayed in figures were estimated by the statistics of data. Here, we like to note that these emission patterns of $K\alpha_1(+M2)$ or $K\alpha_2(+M1)$ transitions of He-like xenon produced by NRDC are clearly different to earlier findings obtained from the Lyman-$\alpha_1$ transition of H-like xenon created by NRC [12,33]. From these figures, one can see that the emission patterns of the $K\alpha_1(+M2)$ transition (normalized to the Lyman-$\alpha_2(+M1)$ intensity) represent significant anisotropy except for the result of 95 MeV/u $Xe^{54+}$+Xe collisions. Moreover, target-atomic-number and energy dependencies of anisotropy parameters were obtained.

However, the angular distributions of the intensity ratios between the $K\alpha_2(+M1)$ and Lyman-$\alpha_2(+M1)$ transitions are nearly isotropic. Here, the contribution of the $^3S_1$ state is isotropic, and the states $^3P_2$ and $^3P_0$ can decay to the ground state via the intermediate state $^3S_1$ [47]. In the present experiments, such events are counted as contributions to the $K\alpha_2(+M1)$ transition. According to nearly isotropy distribution of $K\alpha_2(+M1)$ observed in the present work, the angular emission patterns of the $K\alpha_1(+M2)$ could also be obtained from the angular distribution of the intensity ratio between the $K\alpha_1(+M2)$ and $K\alpha_2(+M1)$ transitions, and can be represented by

$$\frac{I_{K\alpha_1(+M2)}(\theta_{\text{lab}})}{I_{K\alpha_2(+M1)}(\theta_{\text{lab}})} = \frac{\sigma_{K\alpha_1}^{\text{total}}}{\sigma_{K\alpha_2}^{\text{total}}} \times \left\{1 + \beta_{20}^{K\alpha_1(+M2)\prime}\left[1 - \frac{3}{2}\frac{(1-\beta^2)\sin^2\theta_{\text{lab}}}{(1-\beta\cos\theta_{\text{lab}})^2}\right]\right\}. \qquad (7)$$

The corresponding fitting results $\sigma_{K\alpha_1(+M2)}^{\text{total}}/\sigma_{K\alpha_2(+M1)}^{\text{total}}$ and $\beta_{20}^{K\alpha_1(+M2)\prime}$ for normalized $K\alpha_1(+M2)$ by $K\alpha_2(+M1)$ are exhibited in Fig. 5, and then collected in Figs. 6 and 7 for comparison. In additions, the intensity ratios between the $K\alpha$ and Lyman-$\alpha_2(+M1)$ transitions are much smaller than the intensity ratios between the Lyman-$\alpha_1$ and Lyman-$\alpha_2(+M1)$ transitions as presented in previous work [12], which imply that single capture is dominant than double capture, as well as double capture cannot be ignored especially for lower energies and heavier targets. Moreover, the intensities of $K\alpha_2$ are larger than that of $K\alpha_1$ except for 146 MeV/u Xe$^{54+}$+Kr collisions. It should be noted that the experimental results include cascades from higher excited states of down-charged Xe ions.

Surzhykov *et al.* [46,51] reported $\beta_{20}^{K\alpha_1(+M2)}$ results of the $K\alpha_1(+M2)$ transition produced by REC into the H-like xenon ions for the collision energies in the range from 10 to 400 MeV/u. The theoretical $\beta_{20}^{K\alpha_1(+M2)}$ values of the He-like xenon ions are 0.03 and 0.025 for the energies of 95 and 146 MeV/u, respectively. That is to say, the $K\alpha_1(+M2)$ decay gave rise to almost an isotropic emission pattern. Moreover, these anisotropy parameters of $K\alpha_1(+M2)$ radiation from heavy He-like ions following REC by relativistic collisions do not depend on the charge nucleus of the target [51]. The $\beta_{20}^{K\alpha_1(+M2)}$ can be deduced by the alignment parameter of the $2p_{3/2}$ state from simple forms, as well as the alignment parameters of the $[1s_{1/2}, 2p_{3/2}]^1P_1$ and $[1s_{1/2}, 2p_{3/2}]^3P_2$ states. These results represent that anisotropy of the $K\alpha_1$ is always smaller than that of the corresponding Lyman-$\alpha_1$ in the case of the REC process.

In the present work, the obtained $\beta_{20}^{K\alpha_1(+M2)}$ from NRDC process is clearly larger than that of the Lyman-$\alpha_1$ from NRC process except for the result of 95 MeV/u with Xe, the results show strong dependency for the collision energy because the $\beta_{20}^{K\alpha_1(+M2)}$ is from -0.51 at 95 MeV/u to 0.41 at 146 MeV/u for the Kr target. Figs. 6 and 7 plot energy- and target-dependencies of the anisotropy parameters $\beta_{20}^{K\alpha_1(+M2)}$, $\beta_{20}^{K\alpha_2(+M1)}$, and $\beta_{20}^{K\alpha_1(+M2)\prime}$ and the total intensity ratios $\sigma_{K\alpha_1(+M2)}^{\text{total}}/\sigma_{\text{Ly-}\alpha_2(+M1)}^{\text{total}}$, $\sigma_{K\alpha_2(+M1)}^{\text{total}}/\sigma_{\text{Ly-}\alpha_2(+M1)}^{\text{total}}$, and $\sigma_{K\alpha_1(+M2)}^{\text{total}}/\sigma_{K\alpha_2(+M1)}^{\text{total}}$ of projectile Xe$^{52+*}$ produced by the NRDC mechanism, respectively. The previous results of $\beta_{20}^{\text{Ly-}\alpha_1}$ and $\sigma_{\text{Ly-}\alpha_1}^{\text{total}}/\sigma_{\text{Ly-}\alpha_2(+M1)}^{\text{total}}$ of projectile Xe$^{53+*}$ produced by the NRC mechanism are also displayed for comparisons. Clearly, $\beta_{20}^{\text{Ly-}\alpha_1}$ parameters were negative for the energy range considered, but the obtained $\beta_{20}^{K\alpha_1(+M2)}$ and $\beta_{20}^{K\alpha_2(+M1)}$ can take both negative and positive values.

To the best of our knowledge, there is not applicable approach to calculate the magnetic sublevels cross sections (i.e., alignment parameters or anisotropy parameters) of the NRC and NRDC processes in fast collisions of bare high-Z ions with heavy atoms. The experimental findings of this work provide anisotropy parameters of the NRDC in fast collisions of bare highly charged heavy ions for the first time. From the angular-resolved measurement of the x-ray emission following NRDC, we were able to get information of the associated population distributions for the magnetic $L$-shell sublevels belonging to the $^1P_1$, $^3P_2$, $^3P_1$, and $^3S_1$ states in He-like xenon. Compared with the results of NRC into the $2p_{3/2}$ state in H-like xenon [12,33], a very different behavior for the two hydrogen- and helium-like systems is displayed, which may help to reveal the role of the coupling of the electrons and their interaction with nuclei in the high-$Z$ regime.

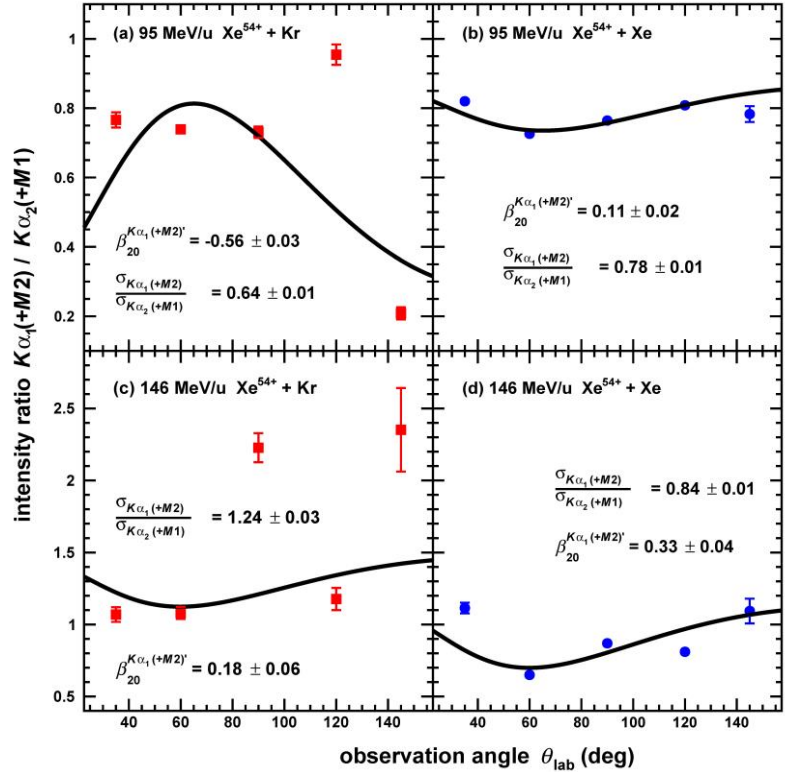

FIG. 5. Angular distributions of the intensity ratios between the $K\alpha_1(+M2)$ and $K\alpha_2(+M1)$ of He-like xenon produced by the nonradiative double electron capture in collisions of 95 and 146 MeV/u Xe$^{54+}$ projectiles with Kr and Xe targets. The solid lines are fittings of the experimental data according to Eq. (7), the anisotropy parameters $\beta_{20}^{K\alpha_1(+M2)\prime}$ and total cross-section ratios $\sigma_{K\alpha_1(+M2)}^{\text{total}}/\sigma_{K\alpha_2(+M1)}^{\text{total}}$ are also displayed. The error bars are estimated by the statistics of data.

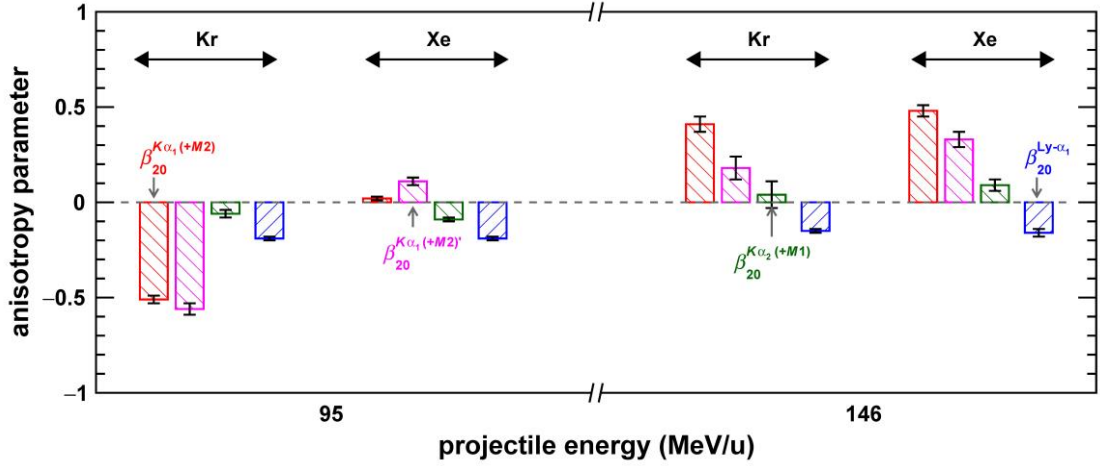

FIG. 6. Anisotropy parameters $\beta_{20}^{K\alpha_1(+M2)}$, $\beta_{20}^{K\alpha_1(+M2)'}$, and $\beta_{20}^{K\alpha_2(+M1)}$ for the $K\alpha_1(+M2)$ and $K\alpha_2(+M1)$ transitions of He-like xenon produced by the nonradiative double electron capture in collisions of 95 and 146 MeV/u Xe$^{54+}$ ions with Kr and Xe atoms. Results of $\beta_{20}^{Ly-\alpha_1}$ for the Lyman-$\alpha_1$ transition of H-like xenon produced by the NRC [12,33] are compared with the experimental findings.

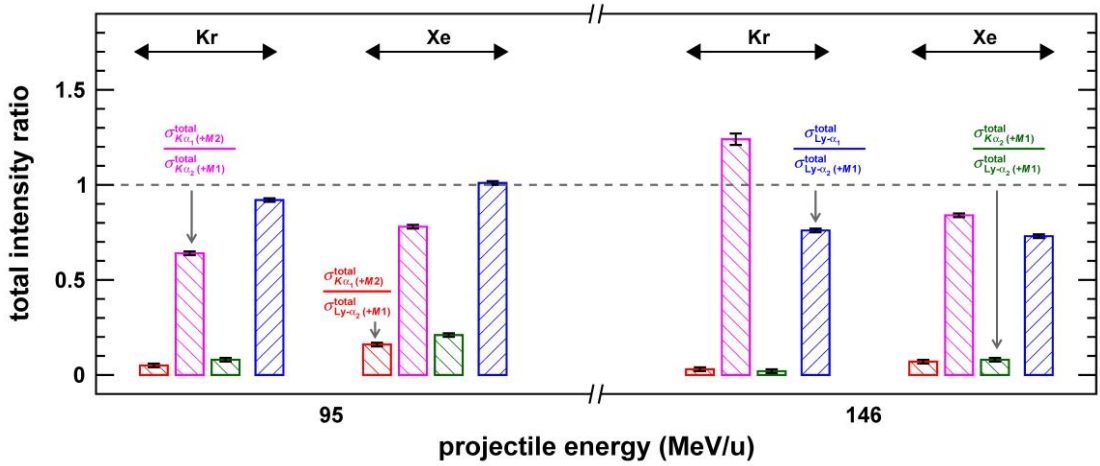

FIG. 7. The total cross-section ratios $\sigma_{K\alpha_1(+M2)}^{total}/\sigma_{Ly-\alpha_2(+M1)}^{total}$, $\sigma_{K\alpha_2(+M1)}^{total}/\sigma_{Ly-\alpha_2(+M1)}^{total}$, and $\sigma_{K\alpha_1(+M2)}^{total}/\sigma_{K\alpha_2(+M1)}^{total}$ for He-like and H-like xenon produced by the nonradiative double and single electron capture in collisions of 95 and 146 MeV/u Xe$^{54+}$ with Kr and Xe atoms. Results of $\sigma_{Ly-\alpha_1}^{total}/\sigma_{Ly-\alpha_2(+M1)}^{total}$ for H-like xenon produced by the NRC [12,33] are compared with the experimental findings.

As seen from Eqs. 3-7, further analysis of the angular distribution of the $K\alpha_1(+M2)$ and $K\alpha_2(+M1)$ radiation can be track back to computations of the alignment as well as the relative populations of the $^1P_1$, $^3P_2$, $^1P_3$, and $^1S_3$ excited ionic states. For NRDC into bare heavy ions, however, the alignment parameters $\mathcal{A}_{20}$ and the weights $N$ are determined not only by the direct electron capture into the particular states but also by the cascade feeding from the high-lying levels. For example, due to a non-ignorable contribution from the $[1s_{1/2}, 2p_{3/2}]\,^3P_2 \to [1s_{1/2}, 2s_{1/2}]\,^3S_1$ decay as shown in Fig. 2, only about 89% of the population of the $^3P_2$ state contributes to the $K\alpha_1$ transition. This branching ratio implies that the wights of the corresponding state and the measured spectral lines (and fine-structure components) in the $K\alpha_1$ transition need to consider these contributions. In fact, such behaviour is quite different from what is known for a strong anisotropy arising for the subsequent Lyman-$\alpha_1$ radiation following NRC. The different (energy) dependence of the anisotropy parameters depicted in Fig. 6, results in qualitatively different emission patterns of the Lyman-$\alpha_1$, $K\alpha_1$(+$M2$), and $K\alpha_2$(+$M1$) radiation following the electron capture by bare xenon ions. In particular, the large negative parameter $\beta_{20}(2p_{3/2} \rightarrow 1s_{1/2})$ leads to a strong angular dependence of the Lyman-$\alpha_1$ radiation whose anisotropy is obviously decreased to nearly zero if one proceeds towards higher collision energies at 197 MeV/u for the Kr target [33]. The $K\alpha_1$(+$M2$) radiation from heliumlike xenon projectiles, in contrast, appears to be the large positive anisotropy parameters for both targets at 146 MeV/u, as the projectile energy decreasing to 95 MeV/u, the anisotropy is decreased to nearly zero for the Xe target, however the anisotropy is changed to much larger negative for the Kr target. Measurements of the $K\alpha_1$(+$M2$) emission allow to obtain composite anisotropy determined by two different transitions (see Eq.(5)). However, without information about relative populations of $[1s_{1/2}, 2p_{3/2}]\,^1P_1$ and $[1s_{1/2}, 2p_{3/2}]\,^3P_2$ states, it is not possible to extract the corresponding alignment parameters. For the case of the $K\alpha_2$(+$M1$) transition, the anisotropy is (negatively) nearly zero for both targets at 95 MeV/u, the values are slightly increased to positive at 146 MeV/u. To be specific, the contribution of the state $[1s_{1/2}, 2s_{1/2}]\,^3S_1$ to the $K\alpha_2$(+$M1$) line is isotropy (almost 88% of population of $[1s_{1/2}, 2p_{1/2}]\,^3P_0$ decays into the $^3S_1$ state), however, the case of the state $[1s_{1/2}, 2p_{1/2}]\,^3P_1$ is essentially depend on the projection of angular momentum [47], but which is proved to almost isotropy in the present work.

From the Fig. 7, one can see that the $K\alpha_1$(+$M2$)/Lyman-$\alpha_2$(+$M1$) intensity ratios for the Kr and Xe target are clearly smaller than the Lyman-$\alpha_1$/Lyman-$\alpha_2$(+$M1$) intensity ratios by ~4%-5% and ~10%-16%. Moreover, the $K\alpha_2$(+$M1$)/Lyman-$\alpha_2$(+$M1$) intensity ratios are separately smaller by ~3%-9% and ~11%-21%. The obtained experimental findings indicate that single electron capture dominates in the present work, and double electron capture is enhanced at the lower energy and heavier target. Moreover, considering the decay scheme of He-like xenon in Fig. 2, the theoretical cross sections for the states of $^1P_1$, $^3P_2$, $^3P_1$, and $^3S_1$ can be calculated for comparison with experimental results by the following formulas

$$\sigma^{total}_{K\alpha_1(+M2)}/\sigma^{total}_{Ly\text{-}\alpha_2(+M1)} = \frac{\sigma(^1P_1)+0.89\sigma(^3P_2)}{\sigma(2s_{1/2})+\sigma(2p_{1/2})}, \qquad (8)$$

$$\sigma^{total}_{K\alpha_2(+M1)}/\sigma^{total}_{Ly\text{-}\alpha_2(+M1)} = \frac{\sigma(^3P_1)+\sigma(^3S_1)+0.11\sigma(^3P_2)+0.88\sigma(^3P_0)}{\sigma(2s_{1/2})+\sigma(2p_{1/2})}, \qquad (9)$$

and

$$\sigma^{total}_{K\alpha_1(+M2)}/\sigma^{total}_{K\alpha_2(+M1)} = \frac{\sigma(^1P_1)+0.89\sigma(^3P_2)}{\sigma(^3P_1)+\sigma(^3S_1)+0.11\sigma(^3P_2)+0.88\sigma(^3P_0)}. \qquad (10)$$

Here, the $\sigma(^{2s+1}L_J)$ refers to the cross-sectional population of the corresponding level and the numbers in front of these cross sections are the branching ratios for the corresponding transitions based on the transitions rates. These formulas include only to $n=2$ states and the subsequent decay. In contrast, the measured experimental results include cascade feeding from higher excited states $n>2$. However, the cascade contributions of higher excited states $n>2$ to the measured intensity ratios turn out to be quite small [17,36].

In the present work, we have performed the first experimental study of the magnetic sublevel population for NRDC of bare xenon in relativistic collisions with heavy gaseous targets. The information about the population of the magnetic sublevels in this process could be obtained via an angular differential study of the decay photons associated with the anisotropy parameters of the corresponding transitions. The results may not be understood within a single-electron model and require one to account for the coupling of the electrons as well as their interaction [17]. Usually, methods of dealing with high-$Z$ ions for which the interaction among the electrons is commonly assumed to be of minor importance. Since this is the only measurement for NRDC into bare heavy ions by relativistic collisions to date, additional studies with different targets, different collision energies, and different ionic species would be desirable to unravel especially the role of electron-nuclei interaction and electron-electron interaction in relativistic collisions of heavy ions with gaseous targets. In addition, more applicable relativistic theories would be highly demanding to quantitatively unravel physical effects behind the NRDC process in relativistic collisions of highly charged high-Z ions with multielectron targets. We hope that the experimental results in the present work could serve as a test for relevant theoretical methods of NRDC mechanism.

## IV. CONCLUSION

In conclusion, we studied experimentally the nonradiative double electron capture in collisions of $Xe^{54+}$ with Kr and Xe atoms at projectile energies of 95 and 146 MeV/u. The anisotropy parameters $\beta_{20}^{K\alpha_1(+M2)}$ and $\beta_{20}^{K\alpha_2(+M1)}$ are obtained by the angular distributions of the intensity ratios $K\alpha_1(+M2)$/Lyman-$\alpha_2(+M1)$ and $K\alpha_2(+M1)$/Lyman-$\alpha_2(+M1)$ from the down-charged projectile ions $Xe^{52+*}$ and $Xe^{53+*}$. The parameters $\beta_{20}^{K\alpha_2(+M1)}$ are nearly isotropy. On the contrary, the $\beta_{20}^{K\alpha_1(+M2)}$ show significantly anisotropy with strong dependency of the collision energy and the target, the corresponding (nonzero) anisotropy parameters include apparently negative and positive values. The features of anisotropy parameters for collision energy and target atomic number show markedly different with that of the Lyman-$\alpha_1$ radiation of $Xe^{53+*}$ produced by the nonradiative single capture. The angular distributions of these radiative transitions behave in qualitatively different ways may be related to the coupling of the electrons to the nuclei and to each other. The obtained results are helpful to reveal the population mechanism of excited states and corresponding magnetic sublevels in collisions of highly charged high-Z ions with multielectron atoms, and might serve as a test (and drive development) for the relativistic theory of NRDC dynamics.

## ACKNOWLEDGMENTS

We thank all participating members of the accelerator department for their operation of the Heavy Ion Research Facility at Lanzhou–Cooling Storage Ring. This work was supported by the National Key Research and Development Program of China under Grant No. 2022YFA1602500; the "Light of West China" Program of Chinese Academy of Sciences under Grant No. xbzglzb2022004; the National Natural

Science Foundation of China under Grants No. 12375263; and Gansu Provincial Talent Young Program under Grant No. 2025QNGR66.## References

[1] D. P. Dewangan and J. Eichler, *Charge exchange in energetic ion-atom collisions*, Phys. Rep. **247**, 60 (1994).

[2] J. Eichler and W. E. Meyerhof, *Relativistic Atomic Collisions* (Academic Press, New York, 1995).

[3] J. Eichler and Th. Stöhlker, *Radiative electron capture in relativistic ion–atom collisions and the photoelectric effect in hydrogen-like high-Z systems*, Phys. Rep. **439**, 1 (2007).

[4] R. Anholt, W. E. Meyerhof, X. Y. Xu *et al.*, *Atomic collisions with relativistic heavy ions. VIII. Charge-state studies of relativistic uranium ions*, Phys. Rev. A **36**, 1586 (1987).

[5] R. Geller, *Electron cyclotron resonance sources: Historical review and future prospects (invited)*, Rev. Sci. Instrum. **69**, 1302 (1998).

[6] R. Becker and O. Kester, *Electron beam ion source and electron beam ion trap (invited)*, Rev. Sci. Instrum. **81**, 02A513 (2010).

[7] P. Micke, S. Kühn, L. Buchauer *et al.*, *The Heidelberg compact electron beam ion traps*, Rev. Sci. Instrum. **89**, 063109 (2018).

[8] M. Steck and Y. A. Litvinov, *Heavy-ion storage rings and their use in precision experiments with highly charged ions*, Prog. Part. Nucl. Phys. **115**, 103811 (2020).

[9] W. L. Zhan, H. S. Xu, G. Q. Xiao *et al.*, *Progress in HIRFL-CSR*, Nucl. Phys. A **834**, 694c (2010).

[10] J. W. Xia, W. L. Zhan, B. W. Wei *et al.*, *The heavy ion cooler-storage-ring project (HIRFL-CSR) at Lanzhou*, Nucl. Instrum. Methods Phys. Res. A **488**, 11 (2002).

[11] X. Cai, R. Lu, C. Shao *et al.*, *Test results of the HIRFL-CSR cluster target*, Nucl. Instrum. Methods Phys. Res. A **555**, 15 (2005).

[12] B. Yang, D. Yu, C. Shao *et al.*, *Alignment of the projectile $2p_{3/2}$ state created by nonradiative electron capture in 95- and 146-MeV/u $Xe^{54+}$ with Kr and Xe collisions*, Phys. Rev. A **102**, 042803 (2020).

[13] F. M. Kröger, G. Weber, M. O. Herdrich *et al.*, *Electron capture of $Xe^{54+}$ in collisions with $H_2$ molecules in the energy range between 5.5 and 30.9 MeV/u*, Phys. Rev. A **102**, 042825 (2020).

[14] X. Ma, Th. Stöhlker, F. Bosch *et al.*, *Subshell differential cross sections for electron transfer in collisions of $U^{90+}$ ions with gaseous targets*, Phys. Scr. **T92**, 362 (2001).

[15] X. Ma, Th. Stöhlker, F. Bosch *et al.*, *State-selective electron capture into He-like $U^{90+}$ ions in collisions with gaseous targets*, Phys. Rev. A **64**, 012704, 012704 (2001).

[16] Th. Stöhlker, T. Ludziejewski, H. Reich *et al.*, *Charge-exchange cross sections and beam lifetimes for stored and decelerated bare uranium ions*, Phys. Rev. A **58**, 2043 (1998).

[17] B. Yang, Z. Wu, D. Yu *et al.*, *Nonradiative double-electron capture in fast collisions of bare $Xe^{54+}$ ions with Kr and Xe gaseous targets*, Phys. Rev. A **110**, 022801 (2024).

[18] W. E. Meyerhoff, H. Gould, C. Munger *et al.*, *Atomic collisions with relativistic heavy-ions. III. Electron capture*, Phys. Rev. A **32**, 3291 (1985).

[19] J. Eichler, *Relativistic eikonal theory of electron capture*, Phys. Rev. A **32**, 112 (1985).

[20] R. Anholt and J. Eichler, *Eikonal calculations of electron-capture by relativistic projectiles*, Phys. Rev. A **31**, 3505 (1985).

[21] N. Toshima and J. Eichler, *Coupled-channels treatment of excitation and charge-transfer in $U^{92+}$ + $U^{91+}$*
**17 / 19**